# Structural changes of $Ti_3SiC_2$ induced by helium irradiation with different doses

Hongliang Zhang[1,3], Ranran Su[1], Liqun Shi[1*], Daryl J O'Connor[2], Haiming Wen[3]

[1] Institute of Modern Physics, Fudan University, Shanghai, China;
[2] School of Mathematical and Physical Sciences, University of Newcastle, Australia,
[3] Department of Materials Science and Engineering, Missouri University of Science and Technology, United States

Email: lqshi@fudan.edu.cn

## Abstract

In this study, the microstructure changes of $Ti_3SiC_2$ MAX phase material induced by helium irradiation and evolution with a sequence of different helium irradiation doses of $5\times10^{15}$, $1\times10^{16}$, $5\times10^{16}$ and $1\times10^{17}$ cm$^{-2}$ at room temperature (RT) were characterized with grazing incidence X-ray diffraction (GIXRD) and Raman spectra analysis. The irradiation damage process of $Ti_3SiC_2$ can be roughly divided into three stages according to the level of helium irradiation dose: (1) for a low damage dose, only crystal and damaged $Ti_3SiC_2$ exit; (2) at a higher irradiation dose, there is some damaged TiC phase additionally; (3) with a much higher irradiation dose, crystal TiC phase could be found inside the samples as well. Moreover, the 450°C $5\times10^{16}$ cm$^{-2}$ helium irradiation on $Ti_3SiC_2$ has confirmed that $Ti_3SiC_2$ has much higher irradiation tolerance at higher temperature, which implies that $Ti_3SiC_2$ could be a potential future structural and fuel coating material working at high temperature environments.

## 1. Introduction

Fusion energy, which is the most possible way to completely solve human's energy problems, faces both theoretical and technological challenges in which the properties of the materials used in the fusion reactor are the most serious[1, 2]. What's more, nuclear fusion power plant designs require materials to perform at temperatures up to over 1000°C [3] in extreme conditions including high flux 14MeV fast neutrons, large amount of evolutionary helium and hydrogen atoms and plasma facing environments[4-7]. Development of new materials which can normally work in extreme environments is one of the most important problems to solve. The MAX phase materials are a group of nano-layered, machinable, ternary carbides or nitrides, in which M is an early transition metal element, A is one of elements in groups 13–16, and X is C or N. More than 60 MAX phase materials have been found or invented so far from more than 600 candidate options. Many of them exhibit exceptional properties[8] and have high radiation damage tolerance due to its M-X ceramic compounds and high damage repairing abilities because of M-A metal structures[9]. These MAX phase materials are attractive candidate materials for structural and fuel coating applications at these extreme conditions[10-12].

Titanium silicon carbide ($Ti_3SiC_2$), which is one of the $M_{n+1}AX_n$ (MAX) phases, was first synthesized in 1960s[13] and regained the researchers' interests a few years ago[14, 15] because of its excellent properties including its degree of ductility[16] and excellent thermal shock resistance[8, 17, 18], while remaining machinable with current conventional tooling[19]. What's more, it has been confirmed that $Ti_3SiC_2$ has great tolerance to radiation due to its unique microstructure [11, 12, 20, 21]. These properties make $Ti_3SiC_2$ MAX phase material ideal for application under extreme radiation conditions, for example, those proposed within future fission and fusion reactors (e.g. GenIV and ITER). A few studies have researched the irradiation tolerance of $Ti_3SiC_2$ MAX phase material using ions including Ni, Au, Xe and Kr [11, 21-23]. These studies have confirmed that $Ti_3SiC_2$ has good damage tolerance to irradiation and remain crystalline even at a very high irradiation dose of 50dpa. What's more, compared with other MAX phase materials such as $Ti_3AlC_2$ and $Ti_2AlC$, $Ti_3SiC_2$ has better irradiation tolerance and larger grain boundary cracking resistance at temperatures above 400°C[24]. Furthermore, it has been confirmed that $Ti_3SiC_2$ is more damage tolerant to heavy ions irradiation in the temperature range 400°C~700 °C which corresponds to the temperature of future reactors[24]. However, most of the irradiation studies of $Ti_3SiC_2$ have been focused on the microstructural changes induced by heavy ion irradiation and few have been carried out on helium irradiation effects which is very important in application of fission and fusion structural materials[25].

In this paper, bulk $Ti_3SiC_2$ MAX phase material samples were irradiated with four helium doses, $5\times10^{15}$, $1\times10^{16}$, $5\times10^{16}$ and $1\times10^{17}cm^{-2}$ at room temperature using 110 keV helium ions. The surface changes are analyzed using Raman spectrum analysis while the microstructure and composition changes are studied with grazing incidence X-ray diffraction (GIXRD) and refined with Rietveld refinement. A model of the helium irradiation damage process for $Ti_3SiC_2$ MAX phase under different damage levels has been established, which follows the sequence that undamaged $Ti_3SiC_2$ will partly change into damaged $Ti_3SiC_2$ at relative low irradiation dose, and then with higher irradiation dose, some of the damaged $Ti_3SiC_2$ will transform into damaged TiC, and further can change into crystal TiC phase at much higher irradiation dose. This process is highly relevant with the helium evolution inside the material. It has also been confirmed that $Ti_3SiC_2$ has much higher irradiation tolerance at higher irradiated temperature.

## 2. Experimental Section

The material used in this work is polycrystalline bulk $Ti_3SiC_2$, prepared by reactive sintering. Stoichiometric mixtures of $3Ti + SiC + C$ were prepared by hand grinding fine Ti (99.9%), SiC (99.9%) and C (graphite, 99.99%) powders under argon followed by cold pressing in a hardened steel die at 180 MPa. The cylindrical samples were sintered under flowing argon gas by heating to 1600°C at 10°C min$^{-1}$, holding for 4 h and returning to room temperature (RT). The powders contained ~2 wt.% Al doping to assist with reactivity. During sintering, a small amount of $Al_2O_3$ was formed in the sample. The as-received specimens were polished with fine metallographic abrasive paper with $Al_2O_3$ suspensions of a size down to 1 μm, cleaned by rinsing in ultrasonic baths of acetone and ethanol, then annealed at 800°C in a vacuum environment of $5 \times 10^{-5}$ Pa for 1 h to release strain.

The $Ti_3SiC_2$ was irradiated with 110 keV $He^+$ incident at 0° to the normal using the tandem accelerator at Institute of Semiconductors, Chinese Academy of Science. The typical irradiation flux was kept below $1.3 \times 10^{13}$ ions cm$^{-2}$s$^{-1}$ to reduce heating due to the $He^+$ ion impact. The irradiation fluence delivered to samples at room temperature were $5 \times 10^{15}$, $1 \times 10^{16}$, $5 \times 10^{16}$ and $1 \times 10^{17}$ ions cm$^{-2}$ and the background pressure during implantation was less than $5 \times 10^{-4}$ Pa for all irradiations. To study the irradiation damage tolerance of $Ti_3SiC_2$ at higher temperature, a higher temperature irradiation was carried out with the same irradiation flux and background pressure at 450°C. The total damage measured in term of displacements per atom was simulated using SRIM-2013[26] with displacement energies of 25, 15 and 28 eV for Ti, Si and C, respectively and an average atomic density for $Ti_3SiC_2$ of $8.34 \times 10^{22}$ atoms/cm$^3$. It is common practice to quote only one level of dpa for an implantation. The damage level obtained from the SRIM-2013 simulation is estimated to be 1 dpa at the surface rising to 5.5 dpa at a depth of 400nm for irradiation fluence of $1 \times 10^{17}$ ionscm$^{-2}$ of 110 keV He

GIXRD data was obtained at beamline BL14B1 of the Shanghai Synchrotron Radiation Facility

at a wavelength of 1.2398Å. BL14B1 is a beamline based on a bending magnet and a Si (111) double crystal monochromator, which was employed to monochromatize the beam. The size of the focus spot was approximately 0.5 mm and the end station was equipped with a Huber 5021 diffractometer. A sodium iodide scintillation detector was used for data collection. The diffraction data were analysed using a Rietveld analysis program (Rietica 7.1). In this paper, we define four different phases in the He irradiated $Ti_3SiC_2$ samples which are $Ti_3SiC_2$ crystal phase, $Ti_3SiC_2$ damaged phase, TiC crystal phase and TiC damaged phase. The $Ti_3SiC_2$ crystal phase was clearly identified from the sharp peaks at the relevant diffraction angles. The $Ti_3SiC_2$ damaged phase identified as the peaks in GIXRD spectra which are usually shortly shifted from the crystal peaks and broadened. The TiC crystal component was clearly identified from the peaks at the appropriate diffraction angles. The TiC damaged phase reflected the quite broad peaks centred at 30° in the scan (fig 4a). This feature has previously been associated with nanocrystalline TiC. The "physical basis" is related to the width. If the peaks are too broad, then they end up contributing to the background. The width of the TiC damage peaks is clear and those of the $Ti_3SiC_2$ damage peaks should not be broader. From the Reitveld refinement, it was possible to identify the different phases in the sample. The uncertainties in the compositions are typically 1-2%.

Raman spectroscopic analysis was performed on a XploRA Laser Raman spectrometer produced by HORIBA JobinYvon. The measurements were conducted using 632.8nm length laser light with a detection range from 100 to 1900$cm^{-1}$ and a total acquisition time of 100s. The spectrum resolution is better than 2$cm^{-1}$. The Raman signal is known to decay exponentially with depth into the surface, with a decay length of approximately 10 nm. The Raman should therefore be sensitive to the surface region only.

## 3. Results

3.1 Grazing incidence XRD patterns and Rietveld Analysis

GIXRD scans of virgin $Ti_3SiC_2$ sample and samples irradiated by 110 keV $He^+$ ions at RT with doses of $5\times10^{15}$, $1\times10^{16}$, $5\times10^{16}$ and $1\times10^{17} cm^{-2}$ were obtained at incidence angles of 0.2°, 0.4°, 0.6° and 1.0°. These angles correspond to X-ray penetration depth of 10nm, 120 nm, 183nm and 305nm, respectively. The penetration depth is the depth over which the X-ray intensity drops to 37% of the intensity at the surface.

Fig.1 shows the GIXRD spectra and the corresponding Rietveld refinement results for the virgin $Ti_3SiC_2$ sample and samples irradiated with $5\times10^{15}cm^{-2}$ $He^+$ at different incidence angles. Comparing with the GIXRD spectra of virgin sample, it is clear that at such a low helium irradiation dose ( about 0.275dpa ), the spectra of the irradiated samples show not only an increase in the background, but also a serious broadening of the peaks which is a strongly indication of a decrease in the crystallinity. The much larger background of the spectra could mostly come from the diffusion scattering due to the randomize Si layers. As calculated before[10, 27], the Si atoms in $Ti_3SiC_2$ MAX phase materials have the lowest migration energy leading to high mobility and the Si layers were strongly disordered while the TiC layers remained mostly unperturbed after irradiation. The broadening of the peaks caused by He irradiation is much larger than that caused by other ions [23, 28] at nearly the same irradiation dose. As could be seen in the spectra, the irradiation induced damage in the surface area where He concentration is low is very serious, which could be confirmed by both the rise of the background and the disappearance of some diffraction peaks, for example peak (006), (101), (102), (103) and (107). With higher incident angles, these peaks appear again especially at the incident angle of 1.0°, corresponding to a depth from the near damage peak region to the surface, which means that at this irradiation dose, $Ti_3SiC_2$ still has good crystal structure. It could be found in the Rietveld refinement results of the $5\times10^{15}cm^{-2}$ irradiated samples that there are two phases, damaged and undamaged $Ti_3SiC_2$ phases. The results show that helium irradiation causes nearly all the crystal $Ti_3SiC_2$ change into damaged phase, especially in the near surface region. The damage decreases with larger incident angles corresponding to the deeper depth.

At a helium irradiation dose of $1\times10^{16}cm^{-2}$, as shown in Fig.2, the GIXRD spectra and the Rietveld refinement results are very similar to that at $5\times10^{15}cm^{-2}$ helium irradiated dose, and there is no notable higher damage comparing with the $5\times10^{15}cm^{-2}$ one. The refinement results also only include the crystal and the damaged $Ti_3SiC_2$ phase. For the near surface region, the damage level is same as the $5\times10^{15}cm^{-2}$ helium irradiated samples, but a slight increasing damage portion occurs at incident angle of 0.6°. These imply that $Ti_3SiC_2$ is tolerant enough to a helium irradiation dose of $1\times10^{16}cm^{-2}$ above which possible formation of TiC phase will take place as indicated in the following.

With a five times higher irradiation dose of $5\times10^{16}cm^{-2}$, the GIXRD spectra and the Rietveld refinement results show different outcomes (Fig. 3). For the 0.2° incident angle spectrum, all the undamaged Ti$_3$SiC$_2$ phase transforms into damaged Ti$_3$SiC$_2$ phase which is very similar with the 5×10$^{15}$cm$^{-2}$ helium irradiated one. At an incident angle of 0.4° and deeper, a new phase which is the damaged TiC phase could be found in the GIXRD spectrum and further quantified by Rietveld analysis. This kind of phase transformation maybe related to the helium bubble evolution inside the sample which will be discussed more specifically in next part. There are three phases in the sample including damaged, undamaged Ti$_3$SiC$_2$ phases and damaged TiC phase. At a higher incident angle of 0.4°, more than 50% of the Ti$_3$SiC$_2$ transforms into a damaged TiC phase which means that the helium irradiation causes a serious phase transformation at this irradiation dose. At the incident angle of 0.6°, the damage component increases further with more than 70% of the Ti$_3$SiC$_2$ changing into damaged TiC phase, and there is only about 20% of the damaged Ti$_3$SiC$_2$ phase with the remaining 10% comprising undamaged Ti$_3$SiC$_2$ phase. When it comes to the incident angle of 1.0°, the damage decreases greatly with only about 10% of the Ti$_3$SiC$_2$ phase changing into damaged TiC phase. There are about 50% of damaged Ti$_3$SiC$_2$ phase and nearly 40% of the undamaged Ti$_3$SiC$_2$ phase. The damage peak is near the incident angle of 0.6°, corresponding to a depth of 183nm.

At a much higher irradiation dose of 1×10$^{17}$cm$^{-2}$, corresponding to a peak damage level of 5.5dpa, the GIXRD spectra and Rietveld refinement results (Fig.4) reveal a much more transformed material. The damage is evident with not only increased background and the notable broaden of the peaks, but also includes a phase transformation from Ti$_3$SiC$_2$ to TiC at depths probed at 1.0° incidence. This is evident by the two new diffraction peaks located at 28.7° and 33.4° corresponding to fcc TiC (111) and (200) which is labeled as crystal TiC phase. These features confirm that the phase transformation is caused principally by the nuclear collision cascades and helium bubble growing during the He implantation. He bubble growth is probably the most important reason for the phase transformation. Further Rietveld refinement results show that with such a high helium irradiation dose of 1×10$^{17}$cm$^{-2}$, in the near surface region, helium irradiation also causes a serious phase transformation from Ti$_3$SiC$_2$ to nano-scale TiC phase which is the damaged TiC shown in the refinement results. More than 60% of the Ti$_3$SiC$_2$ phase transforms into damaged TiC at 0.2° and 0.4° while the remainder is the damaged Ti$_3$SiC$_2$ phase (about 20%) and less than 10% Ti$_3$SiC$_2$ phase. The formation of nano-scale TiC phase is largely attributed to collision cascade effects as described in our previous publications[23, 28], where TiC$_x$ generation is linked to the loss of Si from irradiated Ti$_3$SiC$_2$ followed by the collapse of the Ti$_3$C$_2$ layers to form a defected TiC$_x$ phase. However, the generation of the equivalent amount of decomposed phase TiC$_x$ at much lower He damage dose compared to other ions Si$^+$ or C$^+$ suggests that: (1) the collapse of Ti$_3$SiC$_2$ as a result of the collision is not the sole reason of TiC$_x$ formation; (2) the helium existence prompting the displacement of Si atoms to the sinks such as voids, surface and grain boundaries is possibly another important reason. For phase transformation from Ti$_3$SiC$_2$ to crystal TiC$_x$ occurring in the region of damage peak with high He concentration, another mechanism exists, in which a significant amount of He energetically prefer to stay along the Si plane and this may block the displaced Si atoms from recombining and lead to the decomposition of the MAX phase material through Si element layers. The generation of crystal TiC$_x$ driven by pure He bubble growth has been proved by post annealing after He irradiation which can cause large scale of displacement of Si atoms leading to formation of larger grain size TiC.

3.2 Raman spectra analysis of the irradiated samples

The Raman spectra of the virgin and irradiated samples shown in Fig.5 reveal almost the same damage trends as the GIXRD and the Rietveld results. Helium irradiation causes serious damage to Ti$_3$SiC$_2$ crystal resulting in dramatically broadened Ti$_3$SiC$_2$ Raman peaks. There are six peaks in the spectrum of Ti$_3$SiC$_2$ ternary carbide, as we know from the reference[29], which are at 159, 228, 281, 312, 631 and 678 cm$^{-1}$, where the TiC$_{0.67}$ shows comparable spectra but shifted positions, i.e., 265, 340, 372, 596 and 661 cm$^{-1}$ due to the similarities in their structures. It is plausible to assign that the 627 and 675 cm$^{-1}$ peaks to vibration modes related to the C-Ti-C bonds, the 228, 281, 312 cm$^{-1}$ peaks to vibration modes related to the C-Ti-Si bonds and the sharpness of peaks is related to the order of C atoms in Ti$_3$SiC$_2$. Comparing with the spectrum of the virgin sample, a low dose of helium irradiation (5×10$^{15}$cm$^{-2}$ and 1×10$^{16}$cm$^{-2}$) reveals the disappearance of the peaks at 228, 281cm$^{-1}$ and broadening of the peaks at 596 and 661 cm$^{-1}$ which indicates a serious disorder in the surface region caused by helium irradiation. The background in the high wavenumber region of the spectrum rises a little for the spectra of 5×10$^{15}$cm$^{-2}$ and 1×10$^{16}$cm$^{-2}$ irradiated samples. Increasing the irradiation dose to 5×10$^{16}$cm$^{-2}$, all the peaks occurred in the virgin sample disappear and the Raman spectra show notable peaks around 386 and 590 cm$^{-1}$ related to TiC$_x$ vibrational modes，which is very different from the spectra obtained from other ion irradiated Ti$_3$SiC$_2$ such as C or other ions[29] where no significant Raman peaks from TiC$_x$ crystal phase were evidenced. Moreover, two great broaden peaks appear at around 1335 and 1580 cm$^{-1}$ which are associated with the A$_{1g}$ and the E$_{2g}$ vibrational modes of graphite, indicating that the existence of C which cannot react to form corresponding compound in the samples. Since the height of the peaks at around 1335 and 1580 cm$^{-1}$ is proportion with the dose of helium irradiation which is relevant with the irradiation time, it is very possible that the high graphite peaks are caused by the unreacted C in the material during the helium irradiation induced carbon pollution because of $CO_2$, CO and other organic impurities in the background vacuum system. Since the Raman spectra analysis only detects a surface state, the results are in good agreement with the GIXRD and the Rietveld refinement results that the irradiation induced damage is really serious at the surface region.

## 4. Discussion

The GIXRD spectra and their Rietveld refinement results for $Ti_3SiC_2$ sample irradiated with helium doses from $5\times10^{15}$cm$^{-2}$ to $1\times10^{17}$cm$^{-2}$ at the incident angle of 0.4° (Fig 6) reveal that the He irradiation damage process of $Ti_3SiC_2$ can be roughly divided into three stages according to the helium irradiation dose: (1) a slight damage causing both the expansion of the $Ti_3SiC_2$ unit cell and increase in crystal microstrain; (2) a higher irradiation damage causing the damaged $Ti_3SiC_2$ phase to transform to damaged TiC phase; (3) a much higher irradiation damage causing the a small portion of the damaged $Ti_3SiC_2$ phase to transform to crystal TiC phase.

More specifically, for the first stage, a relatively low dose irradiation only induces the $Ti_3SiC_2$ to change from crystal phase to damaged phase with peak shift due to the changes of the unit cell parameters. The damage is mainly caused by displacement damage from the collision cascades between the helium ions and the atoms of the solid. The implanted He atoms are sufficiently mobile to reach vacancies thus forming He-vacancy complexes and impeding recombination of Frankel pairs. A large concentration of self –interstitial atoms and He interstitial atoms during He implantation can cause a large crystal expansion. Because of the much stronger Ti-C bond compared with the Ti-Si bond[30-32], the He irradiation caused not only the distortion of the Ti-Si bond but also the increase in the *c* parameter. For the second stage, at higher irradiation doses such as $5\times10^{16}$cm$^{-2}$, the helium irradiation causes not only the original crystal structure damage but also the phase transformation from $Ti_3SiC_2$ to damaged TiC phase. With such high irradiation dose, most helium atoms will gather into He clusters along Si plane and cause a high enough helium concentration as the segregation of helium is energetically favorable along the Si layer. The growth of He clusters by absorbing interstitial He atoms is accompanied by the expansion of the lattice space which is accomplished by the ejection of Si atoms and their diffusion to other nearby sinks. Thus, higher helium implantation concentration may lead to the decomposition of the MAX phase at lower temperature into the damaged TiC phase in small volume which corresponds to nanometer scale grain. For the third stage, when the irradiation dose further increases to $1\times10^{17}$cm$^{-2}$, there is small component of crystal TiC phase formed inside the sample. When the irradiation dose rises, the He clusters or small He bubbles should be able to grow by absorbing newly implanted or resolved mobile He atoms and further grow into larger helium bubbles. Thus, the further evolution of the helium bubbles will push out more Si atoms along Si plane and the broken Ti-Si bonds lead to new Ti-C bonds in larger volume which corresponds to the larger grain crystal phase TiC in GIXRD spectra. The Si atoms pushed out from their positions cannot recombine with Si vacancies and migrant to the nearby grain boundaries and then desorbs. It could be understood that when the He dose is high enough, the Si atoms can obtain enough energy to migrant along the Si plane. It is more easily for the Si atoms in the near surface region to migrant to the surface leading to more TiC existing in the near surface region as observed in the GIXRD spectra of $1\times10^{17}$cm$^{-2}$.

High irradiation temperature is considered can greatly help Ti$_3$SiC$_2$ improve the irradiation tolerance of Ti$_3$SiC. To verify this, a high temperature irradiation was performed to test the irradiation tolerance of Ti$_3$SiC$_2$. Fig.7 shows the GIXRD spectra and their Rietveld refinement results for Ti$_3$SiC$_2$ sample irradiated with $5\times10^{16}$cm$^{-2}$ helium ions at 450°C. Comparing with the irradiation sample with same dose but at room temperature, the 450°C irradiation results reveal a much lower damage level where only some amount of the Ti$_3$SiC$_2$ becomes damaged phase and no TiC phase exits in the samples, implying the lowest damage level based on the former hypothesis. What's more, unlike with the room temperature irradiation results in which the damage is the most serious at the He irradiation range area, the irradiation induced damage decreases almost linearly with increasing depth. This phenomenon implies that at higher irradiation temperature, the damage could migrate to the surface slowly and then self-heal. Fig. 8 shows the trend of the Ti$_3$SiC$_2$ *c* and *a* lattice parameters. Compared with the crystal Ti$_3$SiC$_2$, there is a pronounced increase in the basal (c-direction) lattice parameter at high irradiation doses condition, such as $5\times10^{15}$, $1\times10^{16}$, $5\times10^{16}$ and $1\times10^{17}$cm$^{-2}$ for the room temperature irradiation. Only minimal change of the c-LP occurs for the 450°C irradiated samples. At the irradiation dose of

$5\times10^{16}$ cm$^{-2}$, the damaged Ti$_3$SiC$_2$ exhibits the largest c-LP swelling with an increase from 17.94Å to 18.51 Å which is an increase of approximately 3.18%. On the other hand, for the same irradiation dose of $5\times10^{16}$ cm$^{-2}$ but at a higher temperature of 450°C, the c-LP swelling is much smaller, only changing from 17.94 Å to 17.97 Å with an increase of less than 0.2%. The c-LP swelling is even much lower than the $1\times10^{16}$ cm$^{-2}$ room temperature irradiated samples which changes from 17.94Å to 18.08Å with an increase of about 0.9%. The results show that the damage level for the 450°C irradiated sample is much lower than the room temperature irradiated one with the same irradiation dose, which implies Ti$_3$SiC$_2$ has a better irradiation tolerance at high temperature. When it comes to the prismatic (a-direction) lattice parameter (a-LP), the Ti$_3$SiC$_2$ exhibits a slight a-LP reduction at the irradiation doses of $5\times10^{15}$, $1\times10^{16}$ and $5\times10^{16}$ cm$^{-2}$. The damaged Ti$_3$SiC$_2$ shows the most a-LP contraction at $5\times10^{16}$ cm$^{-2}$ for the room temperature irradiated sample, decreasing from 3.067Å to 3.041 Å with a change of 0.85%. However, for the 450°C irradiated samples with the same irradiation dose of $5\times10^{16}$ cm$^{-2}$, the contraction of the a-LP is much slighter, only changing from 3.067Å to 3.062Å, a slight change of 0.16%. The a-LP contraction of the 450°C $5\times10^{16}$ cm$^{-2}$ irradiated samples is even smaller than that of $1\times10^{16}$ cm$^{-2}$, room temperature irradiated samples. The a-LP for the $1\times10^{17}$ cm$^{-2}$ irradiated sample shows a great increase, changing from 3.067Å to 3.139Å with a change of 2.35% which is very different from the lower irradiation doses. The large differences in c-axis and a-axis swelling can result in pronounced strains at randomly oriented grain boundaries.

## 5. Conclusion

The different levels of helium irradiation dose results in Ti$_3$SiC$_2$ show that Ti$_3$SiC$_2$ has good irradiation damage tolerance due to its special microstructure. A low irradiation dose ($5\times10^{15}$ cm$^{-2}$) could cause the microstructure damage of Ti$_3$SiC$_2$ including the expansion of the unit cells, etc. Higher irradiation dose could cause the phase transformation from crystal Ti$_3$SiC$_2$ phase to the damaged and crystal TiC which has high relationship with the helium evolution inside the sample. Ti$_3$SiC$_2$ exhibits different compositions corresponding to different damage doses, which is, for the lowest damage, only crystal and damaged Ti$_3$SiC$_2$; at a higher irradiation dose, there is some damaged TiC phase additionally inside; with a much higher irradiation dose, crystal TiC phase could be found inside the samples. Ti$_3$SiC$_2$ has higher irradiation tolerance at higher irradiated temperature, which implies that

$Ti_3SiC_2$ could be a potential future structural and fuel coating material working at high temperature environments. Further studies are carried on to research the irradiation damage tolerance of $Ti_3SiC_2$ at different He irradiation temperatures.

## 6. Acknowledgment

The authors thank beamline BL14B1 (Shanghai Synchrotron Radiation Facility) for providing the beam time. The authors gratelfully acknowledge the financial support from China Scholarship Council. Our work was also supported by the National Nature Science Foundation of China under grant number 11375046. We acknowledge the financial support of the Australia/China International Linkage program (CH080126) of the Department of Industry, Innovation and Science Research, Australian Government.

**captions**

Figure 1   GIXRD results and composition analysis obtained from Rietveld refinement for $5\times10^{15}$cm$^{-2}$ of He irradiated Ti$_3$SiC$_2$ samples at different incident angles

Figure 2   GIXRD results and composition analysis obtained from Rietveld refinement for $1\times10^{16}$cm$^{-2}$ of He irradiated Ti$_3$SiC$_2$ samples at different incident angles

Figure 3   GIXRD results and composition analysis obtained from Rietveld refinement for $5\times10^{16}$cm$^{-2}$ of He irradiated Ti$_3$SiC$_2$ samples at different incident angles

Figure 4   GIXRD results and composition analysis obtained from Rietveld refinement for $1\times10^{17}$cm$^{-2}$ of He irradiated Ti$_3$SiC$_2$ samples at different incident angles

Figure 5   Raman spectra for different doses of He irradiated Ti$_3$SiC$_2$ samples

Figure 6   GIXRD results and composition analysis obtained from Rietveld refinement for different doses of

He irradiated Ti$_3$SiC$_2$ samples at 0.4° X-ray incidence angle

Figure 7    GIXRD results and composition analysis obtained from Rietveld refinement for 450°C 5×10$^{16}$cm$^{-2}$ of He irradiated Ti$_3$SiC$_2$ samples at different incident angles

Figure 8    Rietveld refinement results for the c-LP(a) and a-LP (b) of crystal and damaged Ti$_3$SiC$_2$ at different irradiation conditions